# Assessing Automated Machine Learning service to detect COVID-19 from X-Ray and CT images: A Real-time Smartphone Application case study

Razib Mustafiz*[1], Khaled Mohsin, MD[2]

[1]School of Computing, Dublin City University, Ireland.
[2]Department of Cardiology, St James's Hospital, Trinity College Dublin, Ireland.

Corresponding Author: *mohammad.mustafiz2@mail.dcu.ie

**ABSTRACT**

*The recent outbreak of SARS COV-2 gave us a unique opportunity to study for a non interventional and sustainable AI solution. Lung disease remains a major healthcare challenge with high morbidity and mortality worldwide. The predominant lung disease was lung cancer. Until recently, the world has witnessed the global pandemic of COVID19, the Novel coronavirus outbreak. We have experienced how viral infection of lung and heart claimed thousands of lives worldwide. With the unprecedented advancement of Artificial Intelligence in recent years, Machine learning can be used to easily detect and classify medical imagery. It is much faster and most of the time more accurate than human radiologists. Once implemented, it is more cost-effective and time-saving. In our study, we evaluated the efficacy of Microsoft Cognitive Service to detect and classify COVID19 induced pneumonia from other Viral/Bacterial pneumonia based on X-Ray and CT images. We wanted to assess the implication and accuracy of the Automated ML-based Rapid Application Development (RAD) environment in the field of Medical Image diagnosis. This study will better equip us to respond with an ML-based diagnostic Decision Support System(DSS) for a Pandemic situation like COVID19. After optimization, the trained network achieved 96.8% Average Precision which was implemented as a Web Application for consumption. However, the same trained network did not perform the same like Web Application when ported to Smartphone for Real-time inference. Which was our main interest of study. The authors believe, there is scope for further study on this issue. One of the main goal of this study was to develop and evaluate the performance of AI-powered Smartphone-based Real-time Application. Facilitating primary diagnostic services in less equipped and understaffed rural healthcare centers of the world with unreliable internet service.*

1. **INTRODUCTION**

In recent years, the availability of machine learning algorithms made available as service has transformed our ability to add AI features to applications. Expertise that was once the realm of hardcore AI experts can now be accessed by a much wider range of developers armed with a cloud subscription. Machine learning was once a domain of data scientists. From 2016 we saw the development of simplified models that gave developers liberty and the ability to create sophisticated Machine Learning models with minimum effort and knowledge. This wizard-like graphical development environment gave non-data scientists to get their hands into experimenting with ML models. Armed with a limited amount of data and transfer learning technology people with no background in data science can construct sophisticated, industry-standard production-ready AI models with the Rapid Application Development(RAD) method. IT giants came forward to the advent of this platform. Microsoft introduced Custom Vision [1], Google introduced Auto ML [2] and Apple introduced Create ML [3] to democratize Machine Learning.

In our experiment, we used Custom Vision. Microsoft Cognitive Services are part of the Microsoft Azure cloud solution. This machine learning tool enables developers to import their images and create computer vision in Microsoft Azure Custom Vision.[1] Custom Vision is built on a pre-trained Convolutional Neural Network (CNN) and facilitates users with a training technique called transfer learning. It enables the creation of AI models using the Custom Vision web application by simply uploading the training images and tagging them into a model builder. Due to transfer learning technology, that starts with a pre-trained model and uses this model as a feature extractor, Custom Vision does not require as many images for training and testing as regular Convolutional Neural Network (CNN). A minimum requirement of 50 images per tag is recommended. It also runs fast even on less powerful computers as the training is done in the cloud.

Models can be created in Custom Vision and run as a web application through a REST API. There is sample code documentation for Curl, C#, Java, Javascript, ObjC, PHP, Python, Ruby in Microsoft Custom Vision documentation site [4]. Custom Vision is the state of the art machine learning technology that supports to export its trained model into Tensorflow, Tensorflowlite, Tensorflowjs, CoreML, ONNX, Dockerfile, and VAIDK format thus creating an opportunity to use it effortlessly in many platforms. Currently, Custom Vision supports Image classification and object detection. In Image classification, it has two subclasses, 1. Multilabel (Multiple tags per image), 2. Multiclass (Single tag per image) in four different domains. These domains are pre-trained so it can be effectively trained with few image samples. Compact domains of the Custom Vision models can be transferred to mobile and edge devices to do real-time on-device inference.

There are two ways to create a Custom Vision project. Visually and programmatically. For this project, we have chosen the graphical development environment for the sake of Rapid Application Development. Anyone can use C# or Python for the same purpose. The codebase is available in the Custom Vision documentation site.

**Objective**

The purpose of our study was to evaluate Microsoft Cognitive Service to detect COVID19 induced pneumonia and ordinary viral or bacterial infection in Lung using X-Ray and CT scan images. We have used Datasets from a recognized and trusted source to build our model. The primary objective is a Smartphone based on device real-time inference system. In this case, the model would run by a mobile device's System on Chip (SoC) and will not require an internet connection for inference with zero latency. This system would be particularly suitable for rural areas of developing countries where internet connection is poor or not available. The secondary solution would be a web portal running the inference through REST API from Custom Vision.

Now, given the nature of The Severe Acute Respiratory Syndrome Coronavirus 2 (SARS-COV-2), which causes respiratory disease as a novel one, the majority of the radiologists are not acquainted enough to detect the virus-related changes from the X-Ray. Moreover, the morphology of COVID-19 and common Pneumonia are hard to differentiate from X-Ray alone without the patient's symptoms by a radiologist.

Here, AI comes into play with the role of an expert assistant. It is much faster and efficient to train a machine over thousands of labeled training data to observe and detect subtle differences between various X-Ray images to train its Artificial Neural Network and classify them quickly which is otherwise not possible by a human eye. A Radiologist can use the app to primarily identify the X-Ray in question and combine it with his/her medical expertise along with the patient's case history before in conjunction with tests like RT PCR/Antibody.

2. RESOURCES AND METHODOLOGY

   2.1 Data collection and preparation

We have collected datasets of COVID19 affected lung X-Ray Images [5], normal and other viruses/bacteria-induced pneumonic lung X-Ray images [6], COVID19 CT Scan and Normal CT scan images [7], Household objects Dataset [8], labeled as Inapplicable in the Model from authenticated, trusted and well-curated sources. Most of the Chest Radiograph Images (CXR) is available in the Poster anterior views (PA). This is a standard chest radiograph referring to the direction of X-Ray beam travel. It is frequently used to aid the diagnosis of acute and chronic conditions in the lungs. A statistics of datasets acquired are given in the table below.

| Dataset Name | Train Data | Test Data | Validation Data |
|---|---|---|---|
| COVID19 X-Ray | 295 | NIL | 25 |
| Normal X-Ray | 1341 | 234 | 8 |
| Pneumonia X-Ray | 3875 | 390 | 8 |
| Normal CT | 397 | NIL | NIL |
| COVID19 CT | 349 | NIL | NIL |
| Household objects | 185 | 123 | NIL |

Table 1: Dataset info used in the Model

Microsoft Cognitive service recommends using at least 50 images of each class to get a better prediction result. It is also recommended that images of all classes should be equal or close to equal for a better performing model. Dissimilarity in image count results in inclined or declined to a particular class while inferring. As it is seen from the table above, the lowest image count dataset is the **COVID19 X-Ray** in terms of medical imagery. So, despite we have more images in other classes, using an image count of more than 295 will result in an imbalanced model. In our experiment, we will see the impact of imbalanced image counts in the model and how does it behave in the real world.

Microsoft Custom Vision doesn't require test data to be supplied separately for performance scoring. Rather it automatically assigns test data from the supplied data for training. So, in our experiment, we used test data for manual performance evaluation. Since those test data are unknown to the model.

### 2.2 Data loading and Model building

To use the Microsoft Cognitive Service, we need to create Custom Vision Training and Prediction resources in Azure. To do so in the Azure portal, we filled out the dialog window on the Create Custom Vision page to create both Training and Prediction resources. We used the same credential to log into the Custom Vision site. Login with the same credential to both Azure and Custom Vision is important as otherwise, it will not be possible to publish the trained model for inference with REST API.

We ran 7 iterations of training in total with different combinations of Data and classes. Here in Custom Vision, images are clustered by tags. The Table below shows 7 different iteration results in a combination of different image counts and Tags.

| Iterations | Tag Name | Precision | Recall | Average Precision | Image Count |
|---|---|---|---|---|---|
| Iteration 7 (Published) | Inapplicable | 100.0% | 100.0% | 100.0% | 275 |
| | Pneumonia | 98.1% | 92.7% | 96.8% | 275 |
| | Covid-19 | 96.5% | 100.0% | 100.0% | 275 |
| | Normal | 94.7% | 98.2% | 99.6% | 275 |
| | Normal CT | 94.1% | 87.3% | 90.0% | 275 |
| | Covid-19 CT | 89.7% | 94.5% | 89.8% | 275 |
| Iteration 6 | Pneumonia | 100.0% | 92.7% | 93.9% | 275 |
| | Inapplicable | 100.0% | 100.0% | 100.0% | 275 |
| | Covid-19 CT | 96.2% | 90.9% | 95.1% | 275 |
| | Normal | 94.8% | 100.0% | 99.0% | 275 |
| | Covid-19 | 94.8% | 100.0% | 99.4% | 275 |
| | Normal CT | 91.1% | 92.7% | 92.1% | 275 |
| Iteration 5 | Pneumonia | 100.0% | 94.9% | 98.5% | 293 |
| | Inapplicable | 100.0% | 100.0% | 100.0% | 171 |
| | Covid-19 | 100.0% | 100.0% | 100.0% | 275 |
| | Normal | 94.9% | 100.0% | 98.6% | 280 |
| | Normal CT | 90.9% | 89.3% | 95.6% | 277 |
| | Covid-19 CT | 89.8% | 91.4% | 94.4% | 289 |
| Iteration 4 | Pneumonia | 98.4% | 90.3% | 96.6% | 1034 |
| | Covid-19 | 98.2% | 98.2% | 98.2% | 275 |
| | Inapplicable | 93.3% | 96.6% | 97.6% | 291 |
| | Negative | 92.6% | 86.2% | 89.7% | 146 |
| | Normal | 90.8% | 99.0% | 97.2% | 1000 |
| | Covid-19 CT | 89.1% | 81.4% | 87.9% | 349 |
| | Normal CT | 84.5% | 88.8% | 91.1% | 397 |
| Iteration 3 (Published) | Inapplicable | 100% | 100% | 100% | 291 |
| | Pneumonia | 97.4% | 92.9% | 96.6% | 1402 |
| | Normal | 93.2% | 92.9% | 96.9% | 1683 |
| | Covid19 | 87.2% | 96.1% | 95.3% | 640 |
| Iteration 2 (Published) | Inapplicable | 100% | 100% | 100% | 291 |
| | Covid19 | 100% | 98.3% | 99.9% | 291 |
| | Pneumonia | 92.1% | 89.2% | 95.3% | 322 |
| | Normal | 88.1% | 92.2% | 97.2% | 317 |
| Iteration 1 (Published) | Covid19 Positive | 100% | 100% | 100% | 25 |
| | Covid19 Negative | 100% | 100% | 100% | 25 |

Table 2: Creating and testing model using 7 iterations.

Custom Vision uses four different pre-trained models in two different domains namely General and Compact. Compact domain lets users download model to be used in real-time in mobile and edge devices. Also, there are two categories of training, Quick and Advanced. Advanced training trains model to detect images with a challenging and fine-grained dataset with poor augmentation setting. For our study, we used advanced training and compact domain from iteration1 to iteration6. Iteration7 is trained in the General domain with advanced training.

### 2.3 Iteration 1

In this iteration, we trained the Microsoft Custom Vision model to detect Covid19 Positive and Covid19 Negative from training dataset X-Ray images. We created two classes of images (25 images each) with the following labels: Covid19 Positive and Covid19 Negative. For the Covid19 Positive label, we used chest X-Ray images of Covid19 Positive patients and for the Covid19 Negative label, we used chest X-Ray images of Normal (Healthy) persons. As it is seen from Table 2, it scored 100% in all three categories of performance measure. However small sample dataset and with only two labels, it is impractical to use in real life.

### 2.4 Iteration 2

In this iteration, we trained the Microsoft Custom Vision model to detect and classify to differentiate Inapplicable, Covid19, Pneumonia, and Normal Chest X-ray images. We created four classes of images with varying counts [See table 2: Iteration 2] with labels: Inapplicable, Covid19, Pneumonia, and Normal. The model was trained on a compact domain with Advanced training for Mobile device inference. We also published the model to obtain the REST API for cloud-based inference. The model did not perform as expected in Mobile inference. However, the Average Precision of the model was 98.6% despite having varying image count.

### 2.5 Iteration 3

In this iteration, we trained the Microsoft Custom Vision model to detect and classify to differentiate Inapplicable, Covid19, Pneumonia, and Normal Chest X-ray images. We created four classes of images with varying counts [See table 2: Iteration 3] with labels: Inapplicable, Covid19, Pneumonia, and Normal. The model was trained on a compact domain with Advanced training for Mobile device inference. We also published the model to obtain the REST API for cloud-based inference. Custom Vision detected imbalanced image data counts and suggested the distribution of image per tag should be uniform to ensure model performance. The model was unstable and we did not port the model for Mobile device inference. However, the Average Precision of the model was 96.6% despite having an imbalanced image data count. [See table 2: Iteration 3]

### 2.6 Iteration 4

In this iteration, we trained the Microsoft Custom Vision model to detect and classify to differentiate Inapplicable, Covid-19, Pneumonia, Normal, Negative, Covid-19 CT, and Normal CT Chest X-ray and CT images. We created seven classes of images with varying counts [See table 2: Iteration 4] with labels: Inapplicable, Covid19, Pneumonia, Normal, Negative, Covid-19 CT, and Normal CT. The model was trained on a compact domain with Advanced training for Mobile device inference. We did not publish the model to obtain REST API for cloud-based inference since Custom Vision detected imbalanced image data counts and suggested the distribution of image per tag should be uniform to ensure model performance. However, the model was ported to Smartphone App to observe and study the behavior of an unstable model for Mobile device inference. Our study showed that the ported Tensorflow model amplified the instability when it comes to mobile device inference. The Average Precision of the model was 95.1% despite having an imbalanced image data count. [See table 2: Iteration 4]

### 2.7 Iteration 5

In this iteration, we trained the Microsoft Custom Vision model to detect and classify to differentiate Inapplicable, Covid-19, Pneumonia, Normal, Covid-19 CT, and Normal CT Chest X-ray and CT images. We created six classes of images with varying counts [See table 2: Iteration 5] with labels: Inapplicable, Covid19, Pneumonia, Normal, Covid-19 CT, and Normal CT. The model was trained on a compact domain with Advanced training for Mobile device inference. We did not publish the model to obtain the REST API for cloud-based inference. However, the model was ported to the Mobile App to observe and study the behavior of an unstable model for Mobile inference. Our study showed that the ported Tensorflow model amplified the instability when it comes to mobile inference. The Average Precision of the model was 98.0% despite having an imbalanced image data count. [See table 2: Iteration 5]

### 2.8 Iteration 6

In this iteration, we trained the Microsoft Custom Vision model to detect and classify to differentiate Inapplicable, Covid-19, Pneumonia, Normal, Covid-19 CT, and Normal CT Chest X-ray and CT images. We created six classes of images with fixed counts (275 images each) [See table 2: Iteration 6] with labels: Inapplicable, Covid19, Pneumonia, Normal, Covid-19 CT, and Normal CT. The model was trained on a compact domain with Advanced training for Mobile device inference. We did not publish the model to obtain the REST API for cloud-based inference. However, the model was ported to the Mobile App to observe and study the behavior of a stable model for Mobile inference. Our study showed that the ported Tensorflow model performs significantly much better when it comes to mobile device inference. We decided to deploy this trained model to build our Smartphone-based diagnostic App. The Average Precision of the model was 96.8% with balanced (275 images each) image data count. [See table 2: Iteration 6]

### 2.9 Iteration 7

Iteration 7 is the same as Iteration 6 except being trained on the General domain with the view to publishing it for REST API consumption. Iteration 7 produces REST API which we used to construct a web application that can be used for inference outside the Custom Vision site. Iteration 7 trained on the General domain exhibits slight improvement over the Compact domain with 97.0% Average Precision [See table 2: Iteration 7].

## 3 Implementation

### 3.1 Web Application

We implemented our web application in PHP programming language and called the REST API in it. Uploaded images are stored in a custom vision site and can be used for training iterations. This raises an issue of data privacy which should be resolved by the respective jurisdiction when implemented. Custom vision permits to download the entire trained network in Tensorflowjs format. Once implemented as a web service, the trained network downloads itself into the browser, and inference is done locally. So no data leaves the browser. This is a more secure and Data protection friendly option. Microsoft Custom Vision has a documentation page on how to consume REST API [9].

### 3.2 Mobile Application

We have downloaded a trained model in TensorflowLite and CoreML format from Custom Vision to build Android and iOS applications. Smartphone applications are capable of real-time inference with zero latency. As the trained model runs entirely on the mobile app, it does not require an internet connection for image classification. This feature particularly important in rural areas of developing countries where internet connection is not reliable or not available at all. Microsoft has a Github repository with a sample code for App building in Android Studio and Apple Xcode [10]. We used Apple MacBook Pro with macOS High Sierra, Intel Core i5, 8GB RAM, and Intel Iris Plus Graphics 655 GPU 1536 MB, 256 SSD to build Android and iOS app. Our Android and iOS applications are publicly available. Mobile App APK file can be downloaded from **https://softavion.com/CXR/download.php**

Github link for the project files:

**https://github.com/razibmustafiz/COVID-19-X-Ray-Detector-Android**
**https://github.com/razibmustafiz/COVID-19-X-Ray-Detector-iOS**

### 4. RESULTS

When testing with two classes of data each one with only 25 images, we achieved 100% Precision, 100% recall, and 100% Average precision for all classes of images. In other words, our AI model could correctly call Covid19 positive and Covid19 negative 100% all the time. As the model was loaded with more classes and images of different counts, we could see the variance in performance metrics. As we experimented with different combinations of data and classes (See Table:2), the most convincing and favorable results were obtained in iteration 6 where we used 275 images for each class (See Table:2). In which we trained the model to detect and classify 6 different classes. After optimizing the model, we obtained a training Precision of 96.1%, Recall accuracy of 96.1%, and Average Precision accuracy of 96.8% (See Table 2). Point to be noted that, we achieved 100% Recall and 99.4% Average Precision for COVID-19 Class. Let A be a true positive, B be a false negative, C is false positive, and D be true negative. Then accuracy, precision, recall, and F1 score can be calculated using the given formula: Accuracy = (A+D)/(A+B+C+D).

The probability threshold *T* controls the trade-off between the true positive rate (i.e., Precision) and true negative rate (i.e., Recall). To investigate the impact of T on the network performance, we performed the inference stage with different values of *T*, including 50%, 40%, 30%, 20%, and 10% for iteration 6. The network performance obtained by our model was reported in Table 3.

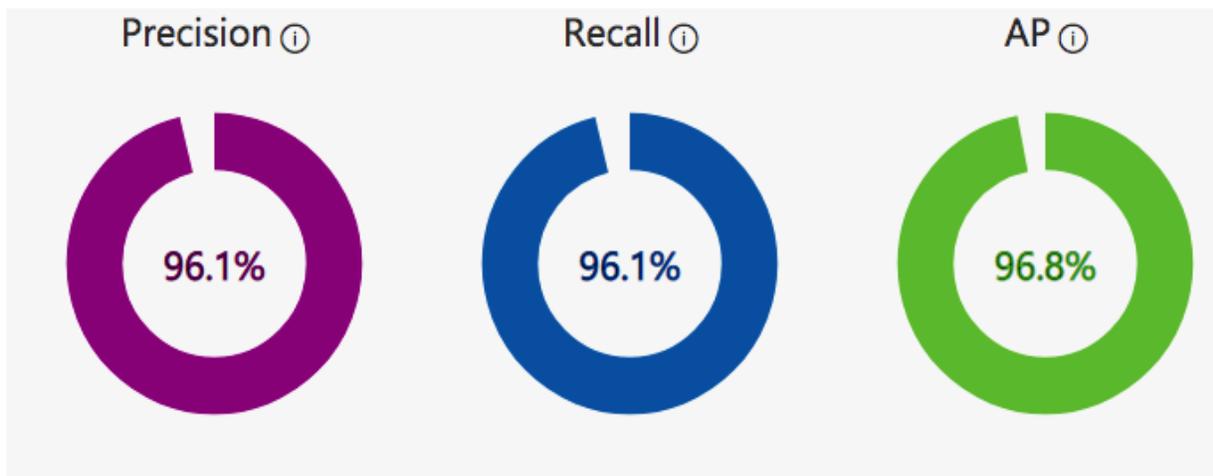

Fig 1: Iteration 6 Performance Metrics in Custom Vision

| Probability Threshold T | Precision % | Recall % | Average Precision % |
|---|---|---|---|
| 50% | 96.1 | 96.1 | 96.8 |
| 40% | 95.8 | 96.1 | 96.8 |
| 30% | 95.5 | 96.4 | 96.8 |
| 20% | 93.0 | 97.3 | 96.8 |
| 10% | 91.2 | 97.6 | 96.8 |

Table 3: Probability Threshold Level *T* in different values.

| Probability Threshold T | 50% | 40% | 30% | 20% | 10% |
|---|---|---|---|---|---|
| Confidence Level (95.0%) | 1.00395230360821 | 1.27476046574023 | 1.65402039763223 | 5.84168632719175 | 8.66248419788644 |
| Upper Confidence Interval (95%) | 97.33728564 | 97.5080938 | 97.88735373 | 101.5416863 | 103.8624842 |
| Lower Confidence Interval (95%) | 95.32938103 | 94.95857287 | 94.57931294 | 89.85831367 | 86.5375158 |

Table 3.1: Confidence Interval for Probability Threshold Level *T*.

Setting the parameter *T* to different values in the inference stage leads interestingly to the same **Average Precision** value, though both the **Precision** and **Recall** are variable and when the parameter *T* decreases from 50% to 10%, the **Precision** drops from 96.1% to 91.2% and the **Recall** increases from 96.1% to 97.6%. As the model primarily trained for COVID-19 detection, the proposed method aims to reduce the false-negative rate as much as possible, since false positive cases can potentially be identified in the subsequent Reverse transcription-polymerase chain reaction (RT-PCR), but false-negative cases will not have a chance for a second test and would be potentially deadly by spreading COVID-19 in their locality. Therefore, we suggest setting the parameter *T* to a small value like 10% to reduce the false-negative rate to as low as 2.4%.

In our study, the Smartphone Application performed poorly. Comparing with web application validation data, the performance matrix is given below:

| Class | Web Application (Average Precision %) | Mobile Application (Average Precision %) |
|---|---|---|
| Pneumonia | 98.1 | 15 |
| Inapplicable | 100 | 100 |
| Covid-19 CT | 89.7 | 40 |
| Normal | 94.7 | 25 |
| Covid-19 | 96.5 | 87 |
| Normal CT | 94.1 | 20 |

Table 4: Performance comparison between Web and Mobile Application.

| Class | Pneumonia | Inapplicable | Covid-19 CT | Normal | Covid-19 | Normal CT |
|---|---|---|---|---|---|---|
| Confidence Level(95.0%) | 527.942806 | 0 | 315.749187 | 442.811235 | 60.354472 | 470.764885 |
| Upper Confidence Interval (95%) | 584.492806 | 100 | 380.599187 | 502.661235 | 152.104472 | 527.814885 |
| Lower Confidence Interval (95%) | -471.392806 | 100 | -250.899187 | -382.961235 | 31.395527 | -413.714885 |

Table 4.1: Confidence Interval for Performance comparison.

We tried to develop a Smartphone-based Real-time COVID-19 detector App with the view to use them in Rural and understaffed areas having access with only an X-Ray machine. The lack of sufficient COVID-19 X-Ray Training Data resulting in poor performance classifying Pneumonic and Normal X-Ray from COVID-19 X-Ray when it comes to Smartphone-based inference. Authors observed that the same trained network performed comparatively poor in Smartphone app than the web app. The magnitude of the performance difference due to Smartphone configuration variance is yet to study.

## 5. DISCUSSION

X-rays are the most common and widely available diagnostic imaging technique, playing a crucial role in clinical care and epidemiological studies [11, 12]. Ordinary care facilities including rural areas have deployed X-ray units as basic diagnostic imaging. Besides, realtime imaging of X-rays with Smartphone Application would significantly speed up the disease screening. Considering these advantages, we aimed to develop a deep learning based model that can detect COVID-19 based on chest X-ray and CT images with adequately high sensitivity with Smartphone-based application. Enabling fast and reliable primary diagnosis. If primary screening is found positive, COVID-19 patients will then be facilitated to be tested with RT-PCR. In this process, a suspected COVID-19 patient will less expose himself to other people. Though X-Ray is cheaper than a CT scan and is more economically viable, detecting COVID-19 using chest X-rays with high sensitivity is very challenging. Not only due to the ribs overlying soft tissue and low contrast but also because of the limited availability of a large number of annotated data. This is specifically true for deep learning-based approaches for image detection and classification. As deep learning is infamously data hungry. Though we have collected a significantly large number of image data of ordinary pneumonia and normal chest X-Ray [6], the unavailability of sufficient COVID-19 induced pneumonic chest X-Ray data prohibited us to use all ordinary pneumonia and normal chest X-Ray data in network training. As this would lead to overfitting and biasing the network. Our training iterations exhibited the result of an imbalanced network (See Table:2). To address the data imbalance problem, we can synthetically generate new COVID-19 X-Ray image data from existing data by using a special type of Neural network called Generative Adversarial Network(GAN).

As we used monotonous training image data with only Anterior-Posterior (AP) and Posterior-Anterior (PA) position, the authors think High counts of augmented training data can solve the problem. As in Inapplicable class, we could see a high level of augmented data is available and the model achieved 100% accuracy with the same 275 image data count. In that view, we can employ the Generative Adversarial Network (GAN) to generate new augmented data from existing data to improve the Smartphone inference accuracy of COVID-19 X-Ray. We have tested six Android Smartphone with different configuration running our app and found out that, newer Smartphone issued after 2018 with faster Processor, high megapixel camera module and bigger RAM performs better than others. This is because; the trained Tensorflow neural network runs on the mobile device's processor for real-time image classification. We did not test the iPhone App in our study.

In medical imaging, it remains a challenging goal on how to generate realistic medical images completely different from the original ones. Synthetic images obtained from Generative Adversarial Network (GAN) would improve diagnostic reliability. Allowing for data augmentation in computer-assisted diagnosis where real data are in scarcity. A plain vanilla GAN, first coined by Goodfellow et al., 2014[20] is a generative model that was designed for directly drawing samples from the desired data distribution without the need to explicitly model the underlying probability density function. It consists of two neural networks: the generator G and the discriminator D. The input to G, z is pure random noise sampled from a prior distribution p(z), which is commonly chosen to be a Gaussian or a uniform distribution for simplicity. The output of G, $X_g$ is expected to have visual similarity with the real sample $X_r$ that is drawn from the real data distribution $P_r(x)$. We denote the nonlinear mapping function learned by G parameterized by $\theta_g$ as $X_g = G(z; \theta_g)$. The input to D is either a real or generated sample. The output of D, $y_1$ is a single value indicating the probability of the input being a real or fake sample. The mapping learned by D parameterized by $\theta_d$ is denoted as $y_1 = D(x; \theta_d)$.

The generated samples form a distribution Pg(x) which is desired to be an approximation of Pr(x) after successful training. D's objective is to differentiate these two groups of images whereas the generator G is trained to confuse the discriminator D as much as possible. Intuitively, G could be viewed as a forger trying to produce some quality counterfeit material, and D could be regarded as the policeman trying to detect the forged items.

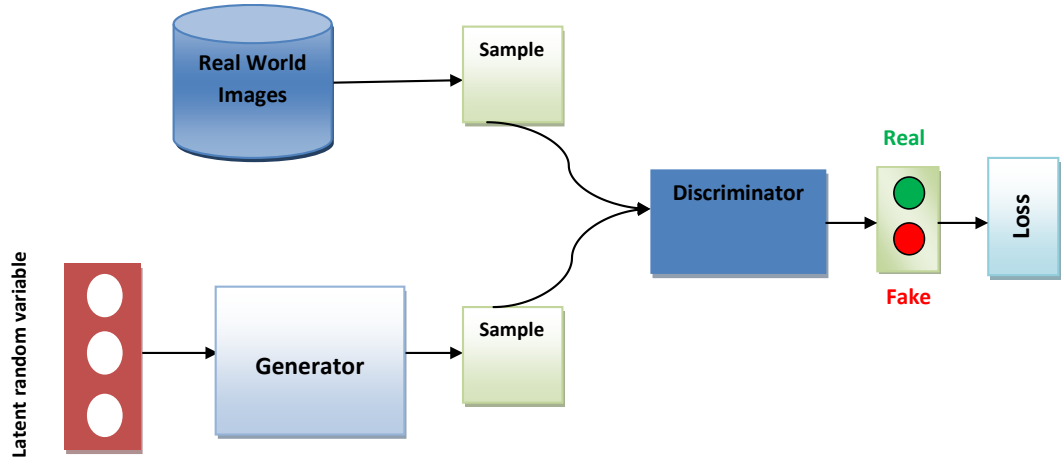

Fig 2: A Plain GAN Schematic Diagram

In an alternative view, we can perceive G as receiving a reward signal from D depending upon whether the generated data is accurate or not. The gradient information is back-propagated from D to G, so G adapts its parameters to produce an output image that can fool D. The training objectives of D and G can be expressed mathematically as:

$$L_D^{GAN} = \max_D E_{x_r \sim p_r(x)}[\log D(x_r)] + E_{x_g \sim p_g(x)}[\log(1 - D(x_g))]$$

$$L_G^{GAN} = \min_G E_{x_g \sim p_g(x)}[\log(1 - D(x_g))].$$

As can be seen, D is simply a binary classifier with a maximum log-likelihood objective. If the discriminator D is trained to optimality before the next generator G updates, then minimizing $L_G^{GAN}$ is proven to be equivalent to minimizing the Jensen Shannon (JS) divergence between *Pr(x)* and *Pg(x)*. The desired outcome after training is that samples formed by $x_g$ should approximate the real data distribution *Pr(x)*.

The inception and advancement of machine learning programs are rapidly changing many aspects of health care. Various studies involving artificial intelligence have been performed in the areas of dermatology, ophthalmology, radiology, and pathology [31,32]. AI has also been utilized in the classification and detection of infectious diseases as well as cardiology programs that assist in identifying patients with heart failure, improving cardiovascular risk predictions, and improving heart failure survival analysis.

The performance advantages of AI programs will be essential in worldwide modern healthcare. Recent AI studies, attempted with various Smartphone applications,[33,34,35] have experimented with microscopy and diagnosis of dermatology lesions[37]. Data analysis composed of diagnostic images, genetic expression testing, and electrophysiological procedures, is converted into valuable assets which may be utilized in treatment decisions, thus reducing errors and improving overall outcomes.

It has been shown in previous studies that, AI programs demonstrated successful learning ability from large numbers of data sets. Providing further classification into subcategories that are then more easily diagnosed and interpreted. Performed on large data sets, these pioneer studies provided further inspiration for useful deep learning studies on even smaller sets of X-Ray clinical images.

In this study, we investigated the ability of an automated AI service to detect and classify COVID-19 induced pneumonia in a Smartphone-based realtime application from a limited number of dataset. Our study primarily reveals that; properly augmented dataset has a huge impact on Smartphone-based real-time inference. A Neural Network trained on a properly augmented dataset can be used in various AI Android/iPhone applications. Smartphone-based diagnosis technologies like this will become progressively significant and vital in remote and rural areas of the developing world lacking accessible healthcare facilities.

### Limitations and further study

The study is a retrospective one based on publicly available datasets. A prospective study based on demographically and morphologically distinct patients would be more relevant. The paucity of published data on ethnically equivalent population made comparison impossible. Further work can be done by generating synthetic COVID-19 X-Ray data with Deep Convolutional Generative Adversarial Networks (DCGAN) and the Progressive Growing Generative Adversarial Networks (PGGAN) for better augmentation in model training. Research [67] showed that PGGAN can produce a better image than DCGAN. However, this raises the question of ethical and legal issues in real-life implementation. But we can use it in our realtime Smartphone App for performance comparison. This study may pave the way for further research and development in this intriguing field.

## 6. CONCLUSIONS

Our study shows how an automated AI service can be used to rapidly build and deploy a medical diagnostic service that can do primary screening in a pandemic situation like COVID-19. Recent advances in AI technology democratize AI to be used in various industries. Smartphone-based imaging and sensing platforms are emerging as promising alternatives for bridging the gap and decentralizing diagnostic tests offering practical features such as portability, cost-effectiveness, and connectivity. In 2020, the current global population is 7.7 billion and the Smartphone penetration rate is at 45.4 percent. In other words, more than four out of every ten people in the world are currently equipped with a Smartphone. A Smartphone-based primary diagnostic service can bring a huge difference in the quality of life, especially in developing countries.

**Declaration of Conflicting Interests**
The authors declare that there is no conflict of interest regarding the publication of this article.
Authors received no financial support for the research, authorship, and/or publication of this article.

**ACKNOWLEDGEMENTS**
This work has not received any kind of funding.

**REFERENCES**

1. Microsoft_Corp. Microsoft Custom Vision. Secondary Microsoft Custom Vision. https://azure.microsoft.com/en-us/services/cognitive-services/custom-vision-service/. Accessed May2020.
2. Alphabet_Inc. Cloud AutoML. Secondary Cloud AutoML. https://cloud.google.com/automl/. Accessed May 2020.
3. Apple_Inc. Create ML. Secondary Create ML. https://developer.apple.com/documentation/createml. Accessed May 2020.
4. Microsoft_Corp. Microsoft Custom Vision. https://docs.microsoft.com/en-us/azure/cognitive-services/custom-vision-service/use-prediction-api . Accessed May 2020
5. Joseph Paul Cohen, Paul Morrison, Lan Dao et al. COVID-19 image data collection, arXiv:2003.11597, 2020https://github.com/ieee8023/covid-chestxray-dataset. Accessed May 2020
6. Chest X-Ray images (Pneumonia) by Paul Timothy Mooney, Developer Advocate at Kaggle.com.https://www.kaggle.com/paultimothymooney/chest-xray-pneumonia. Accessed May 2020
7. Italian society of Medical and Interventional Radiology. https://www.sirm.org/en/2020/03/31/covid-19-case-4/. Accessed May 2020
8. Caltech Home Objects Dataset.http://www.vision.caltech.edu/pmoreels/Datasets/Home_Objects_06/. Accessed May 2020
9. Use your model with the prediction API. https://docs.microsoft.com/en-us/azure/cognitive-services/custom-vision-service/use-prediction-api. Accessed May 2020
10. Sample iOS application for CoreML models exported from Custom Vision Service. https://github.com/Azure-Samples/cognitive-services-ios-customvision-sample . Accessed 2020
11. Thomas Cherian, E Kim Mulholland, John B Carlin et al. Standardized interpretation of paediatric chest radiographs for the diagnosis of pneumonia in epidemiological studies. Bulletin of the World Health Organization, 83:353–359, 2005.
12. T Franquet. Imaging of pneumonia: trends and algorithms. European Respiratory Journal, 18(1):196–208, 2001.
13. C. Dallet, S. Kareem and I. Kale, "Real time blood image processing application for malaria diagnosis using mobile phones," 2014 IEEE International Symposium on Circuits and Systems (ISCAS), Melbourne VIC, 2014, pp. 2405-2408, doi: 10.1109/ISCAS.2014.6865657.
14. D. Jayatilake, K.Suzuki,Y.Teramoto et al., "Swallowscope: A smartphone based device for the assessment of swallowing ability," IEEE-EMBS International Conference on Biomedical and Health Informatics (BHI), Valencia, 2014, pp. 697-700, doi: 10.1109/BHI.2014.6864459.
15. J. J. Oresko, Z.Jin,J.Cheng et al. "A Wearable Smartphone-Based Platform for Real-Time Cardiovascular Disease Detection Via Electrocardiogram Processing," in IEEE Transactions on Information Technology in Biomedicine, vol. 14, no. 3, pp. 734-740, May 2010, doi: 10.1109/TITB.2010.2047865.
16. Michael G. Mauk, Calling in the test: Smartphone-based urinary sepsis diagnostics, https://doi.org/10.1016/j.ebiom.2018.09.001.
17. I Hernández-Neuta , F Neumann , J Brightmeyer et al. Smartphone-based clinical diagnostics: Towards democratization of evidence-based health care

18. A.Borkowski,N.A Viswanadham,L.B Thomas et al. Using Artificial Intelligence for COVID-19 Chest X-ray Diagnosis, doi:10.1101/2020.05.21.20106518.
19. Xin Yi, Ekta Walia, Paul Babyn. Generative Adversarial Network in Medical Imaging: A Review.
20. Ian J. Goodfellow, Jean Pouget-Abadie, Mehdi Mirza et al. Generative Adversarial Nets.
21. C.Han,H.Hayashi,L.Rundo et al. "GAN-based synthetic brain MR image generation," 2018 IEEE 15th International Symposium on Biomedical Imaging (ISBI 2018), Washington, DC, 2018, pp. 734-738, doi: 10.1109/ISBI.2018.8363678.
22. RSNA Pneumonia Detection Challenge | Kaggle. Accessed May 07, 2020. https://www.kaggle.com/c/rsna-pneumonia-detection- challenge .
23. Bloice MD, Roth PM, Holzinger A. Biomedical image augmentation using Augmentor. Murphy R, ed. Bioinformatics. 2019;35(21):4522-4524. doi:10.1093/bioinformatics/btz259 .
24. Cepheid | Xpert® Xpress SARS-CoV-2 has received FDA Emergency Use Authorization. https://www.cepheid.com/coronavirus . Accessed May 10, 2020.
25. Yuen Frank Wong H, Yin Sonia Lam H, Ho-Tung Fong A, et al. Frequency and Distribution of Chest Radiographic Findings in COVID-19 Positive Patients Authors.
26. Fang Y, Zhang H, Xie J, et al. Sensitivity of Chest CT for COVID- 19: Comparison to RT-PCR. doi:10.1016/S0140-6736(20)30211-7
27. Lomoro P, Verde F, Zerboni F, et al. COVID-19 pneumonia manifestations at the admission on chest ultrasound, radiographs, and CT: single-center study and comprehensive radiologic literature review. Published online 2020. doi:10.1016/j.ejro.2020.100231
28. Salehi S, Abedi A, Balakrishnan S, Gholamrezanezhad A. Coronavirus disease 2019 (COVID-19) imaging reporting and data system (COVID-RADS) and common lexicon: a proposal based on the imaging data of 37 studies. doi:10.1007/s00330-020-06863-0
29. Jacobi A, Chung M, Bernheim A, Eber C. Cardiothoracic Imaging Portable chest X-ray in coronavirus disease-19 (COVID-19): A pictorial review. Published online 2020. doi:10.1016/j.clinimag.2020.04.001
30. Bhat R, Hamid A, Kunin JR, et al. Chest Imaging in Patients Hospitalized With COVID-19 Infection - A Case Series. Curr Probl Diagn Radiol. Published online 2020. doi:10.1067/j.cpradiol.2020.04.001
31. Daniel Kermany AS, Goldbaum M, Cai W, Anthony Lewis M, Xia H, Zhang Correspondence K. Identifying Medical Diagnoses and Treatable Diseases by Image-Based Deep Learning. Cell. 2018;172:1122-1131.e9. doi:10.1016/j.cell.2018.02.010
32. Borkowski AA, Wilson CP, Borkowski SA et al. Comparing Artificial Intelligence Platforms for Histopathologic Cancer Diagnosis. Fed Pract. 2019;36(10):456-463.
33. West D. How mobile devices are transforming healthcare. Issues in technology innovation 2012; 18: 1-11.
34. Ozcan A. Mobile phones democratize and cultivate next-generation imaging, diagnostics and measurement tools. Lab Chip 2014; 14: 3187-94.
35. Steinhubl SR, Muse ED, Topol EJ. The emerging field of mobile health. Sci Transl Med 2015; 7: 283rv3.
36. Mudanyali O, Dimitrov S, Sikora U, Padmanabhan S, Navruz I, Ozcan A. Integrated rapid-diagnostic-test reader platform on a cellphone. Lab Chip 2012; 12: 2678-86.
37. Contreras-Naranjo JC, Wei QS, Ozcan A. Mobile Phone-Based Microscopy, Sensing, and Diagnostics. Ieee J Sel Top Quant 2016; 22.
38. Vashist SK, Mudanyali O, Schneider EM, Zengerle R, Ozcan A. Cellphone-based devices for bioanalytical sciences. Anal Bioanal Chem 2014; 406: 3263-77.
39. Orth A, Wilson ER, Thompson JG, Gibson BC. A dual-mode mobile phone microscope using the onboard camera flash and ambient light. Sci Rep 2018; 8: 3298.
40. Smith ZJ, Chu K, Espenson AR et al. Cell-phone-based platform for biomedical device

development and education applications. PLoS One 2011; 6: e17150.
41. Rivenson Y, Ceylan Koydemir H, Wang H, et al. Deep learning enhanced mobile- phone microscopy. ACS Photonics 2018; 5: 2354-64.
42. Yang Z, Zhan Q. Single-Shot Smartphone-Based Quantitative Phase Imaging Using a Distorted Grating. PloS one 2016; 11: e0159596.
43. Martinez AW, Phillips ST, Carrilho E, Thomas SW, 3rd, Sindi H, Whitesides GM. Simple telemedicine for developing regions: camera phones and paper-based microfluidic devices for real-time, off-site diagnosis. Anal Chem 2008; 80: 3699-707.
44. Lipowicz M, Garcia A. Handheld device adapted to smartphone cameras for the measurement of sodium ion concentrations at saliva-relevant levels via fluorescence. Bioengineering 2015; 2: 122-38.
45. Gallegos D, Long KD, Yu H, et al. Label-free biodetection using a smartphone. Lab Chip 2013; 13: 2124-32.
46. Jianpeng Zhang, Yutong Xie, Yi Li et al.COVID-19 Screening on Chest X-ray Images Using Deep Learning based Anomaly Detection.
47. Yan Bai, Lingsheng Yao, Tao Wei, Fei Tian, Dong-Yan Jin, Lijuan Chen, and Meiyun Wang. Presumed asymptomatic carrier transmission of COVID-19. Journal of the American Medical Association (JAMA), 2020.
48. Yicheng Fang, Huangqi Zhang, Jicheng Xie, Minjie Lin, Lingjun Ying, Peipei Pang, and Wenbin Ji. Sensitivity of chest CT for COVID-19: comparison to RT-PCR. Radiol- ogy, page 200432, 2020.
49. Ophir Gozes, Maayan Frid-Adar, Hayit Greenspan, Patrick D Browning, Huangqi Zhang, Wenbin Ji, Adam Bernheim, and Eliot Siegel. Rapid AI development cycle for the coronavirus (COVID-19) pandemic: Initial results for automated detection & patient monitoring using deep learn- ing CT image analysis. arXiv preprint arXiv:2003.05037, 2020.
50. Kaiming He, Xiangyu Zhang, Shaoqing Ren, and Jian Sun. Deep residual learning for image recognition.
51. Feng Shi, Liming Xia, Fei Shan, Dijia Wu, Ying Wei, Huan Yuan, Huiting Jiang, Yaozong Gao, He Sui, and Dinggang Shen. Large-scale screening of covid-19 from community acquired pneumonia using infection size-aware classifica- tion, 2020.
52. Han SS, Park GH, Lim W, et al. Deep neural networks show an equivalent and often superior performance to dermatologists in onychomycosis diagnosis: Automatic construction of onychomycosis datasets by region-based convolutional deep neural network.
53. LiY,ShenL.Skin Lesion Analysis towards Melanoma Detection Using Deep Learning Network.
54. MahbodA,EckerR,EllingerI.Skin Lesion Classification Using Hybrid Deep Neural Networks. arXiv:1702.08434v1 [cs.CV] 27 Feb2017.
55. De Fauw J, Ledsam JR, Romera-Paredes B, et al. Clinically applicable deep learning for diagnosis and referral in retinal disease. Nat Med 2018;24(9):1342-1350.
56. Langs G, Röhrich S, Hofmanninger J, et al. Machine learning: from radiomics to discovery and routine. Radiologe 2018 doi: 10.1007/s00117-018-0407-3[Epub ahead ofprint].
57. McBee MP, Awan OA, Colucci AT, et al. Deep Learning in Radiology. Acad Radiol 2018doi: 10.1016/j.acra.2018.02.018[Epub ahead ofprint].
58. Monkam P, Qi S, Xu M, et al. CNN models discriminating between pulmonary micro-nodules and non-nodules from CT images. Biomed Eng Online 2018;17(1):96.
59. Syeda-Mahmood, Tanveer. (2018). Role of Big Data and Machine Learning in Diagnostic Decision Support in Radiology. Journal of the American College of Radiology.
60. Thrall JH, Li X, Li Q, et al. Artificial Intelligence and Machine Learning in Radiology: Opportunities, Challenges, Pitfalls, and Criteria for Success. J Am Coll Radiol 2018.
61. Yue W, Wang Z, Chen H, et al. Machine Learning with Applications in Breast Cancer Diagnosis and Prognosis. Designs 2018;2(2):13.

62. Poostchi M, Silamut K, Maude RJ, et al. Image analysis and machine learning for detecting malaria.
63. Panahiazar M, Taslimitehrani V, Pereira N, et al. Using EHRs and Machine Learning for Heart Failure Survival Analysis.
64. Abbott LM, Smith SD. Smartphone apps for skin cancer diagnosis: Implications for patients and practitioners. Australas J Dermatol 2018;59(3):168-170.
65. Diederich B, Wartmann R, Schadwinkel H, et al. Using machine-learning to optimize phase contrast in a low-cost cellphone microscope. PLoS One 2018;13(3):e0192937.
66. Wilson ML, Fleming KA, Kuti MA, et al. Access to pathology and laboratory medicine services: a crucial gap. Lancet 2018;391(10133):1927-38.
67. Vassili Kovalev ,Siarhei Kazlouski. Examining the Capability of GANs to Replace Real Biomedical Images in Classification Models Training
68. JiangF,JiangY,ZhiH,et al. Artificial intelligence in health care: past, present and future. Stroke Vasc Neurol 2017;2(4):230-43.
69. Karras T., Aila T., Laine S., Lehtinen J.: Progressive Growing of GANs for Improved Quality, Stability, and Variation. arXiv preprint arXiv:1710.10196 (2017).
70. Radford A., Metz L., Chintala S.: Unsupervised representation learning with deep convolutional generative adversarial networks. arXiv preprint arXiv:1511.06434 (2016).
71. Xin Yia, Ekta Waliaa, Paul Babyna.: Generative Adversarial Network in Medical Imaging: A Review. arXiv preprint arXiv:1809.07294 (2019).
72. Goodfellow I., Pouget-Abadie J., Mirza M., Xu B., Warde-Farley D., Ozair S., Courville A., Bengio Y.: Generative Adversarial Nets. In: Z.Ghahramani et al. (Eds), Neural Information Processing Systems (NIPS) 2014, vol. 2, pp. 2672–2680. MIT Press, Montreal (2014).
73. Salome Kazeminiaa, Christoph Baurb, Arjan Kuijperc, Bram van Ginnekend, Nassir Navabb, Shadi Albarqounib, Anirban Mukhopadhyaya: GANs for Medical Image Analysis. arXiv preprint arXiv:1809.06222 (2018).
74. Kovalev V., Liauchuk V., Kalinovsky A., Shukelovich A. A.: Comparison of conventional and Deep Learning methods of image classification on a database of chest radiographs. International Journal of Computer Assisted Radiology and Surgery 12, 139-140 (2017).

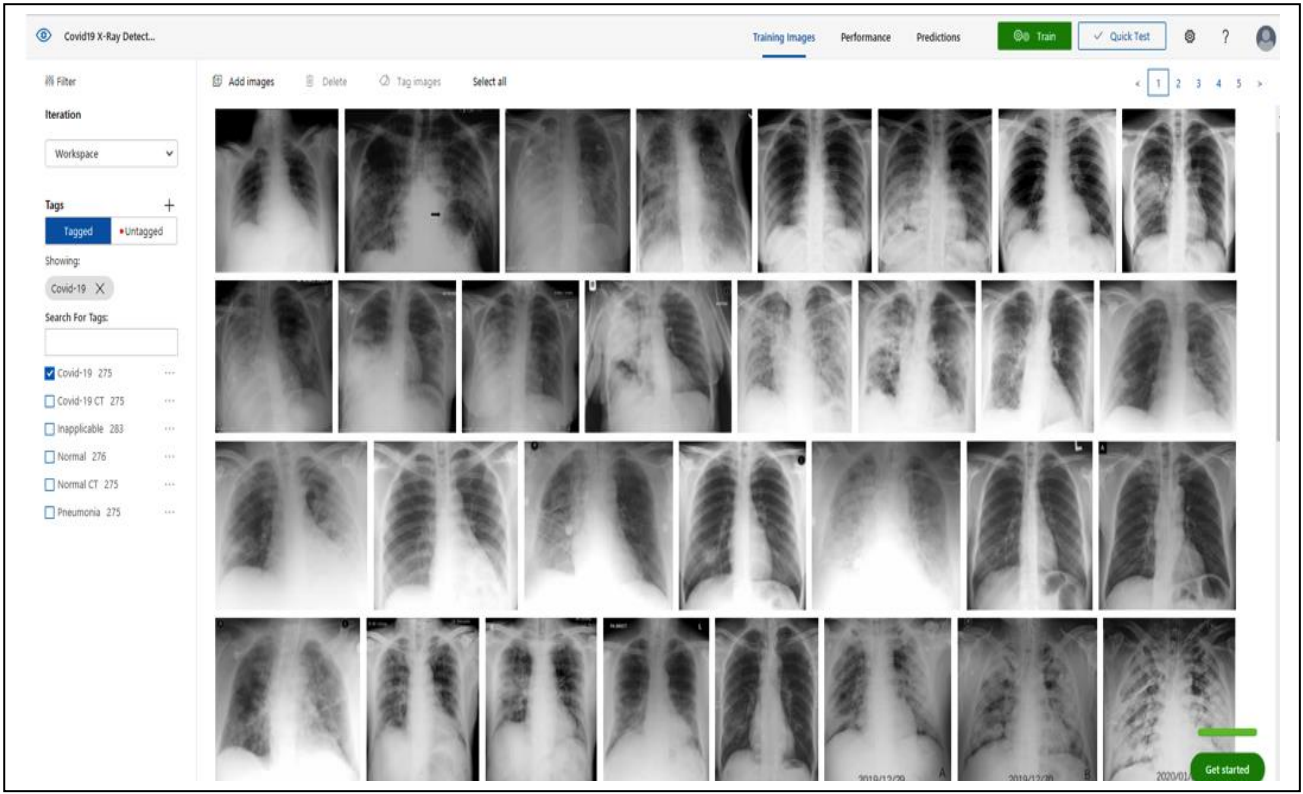

Fig 3: COVID-19 X-Ray Data loaded in Custom Vision

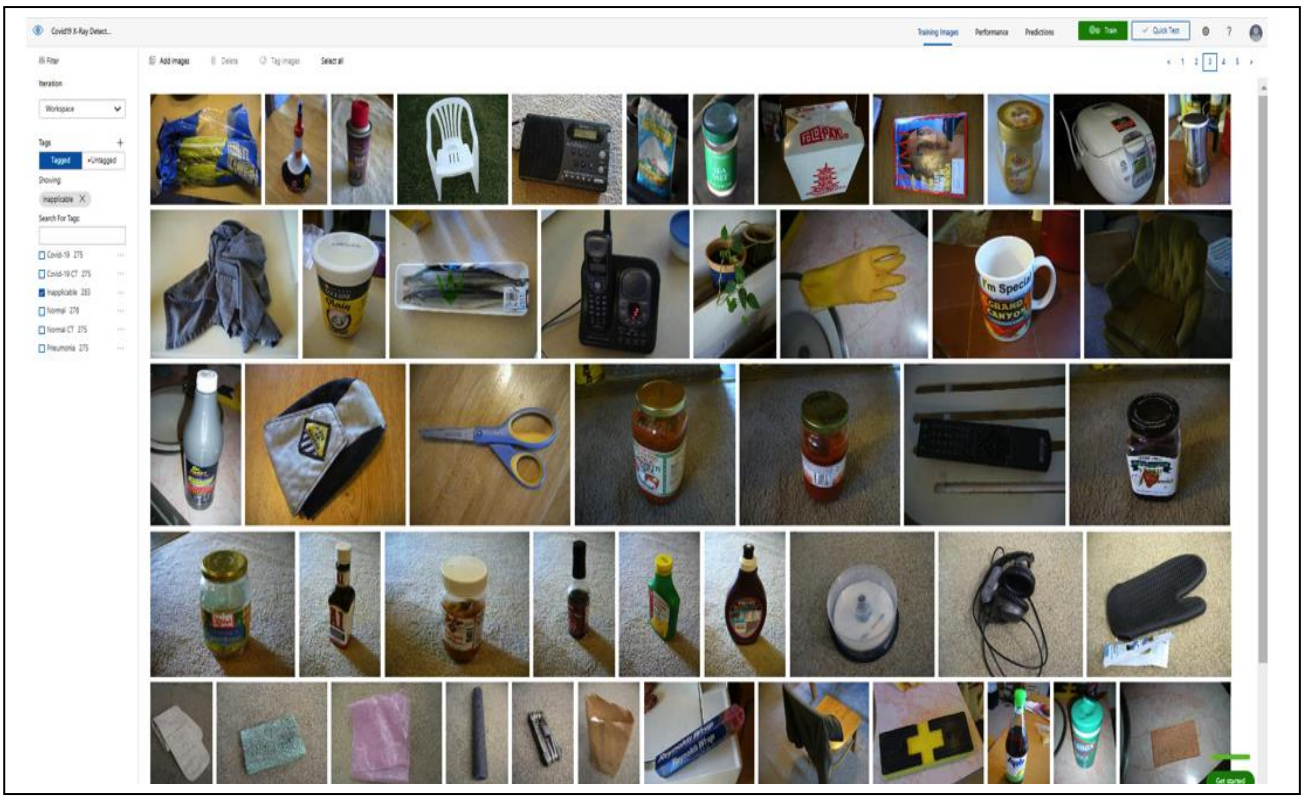

Fig 4: Household Objects Data loaded in Custom Vision

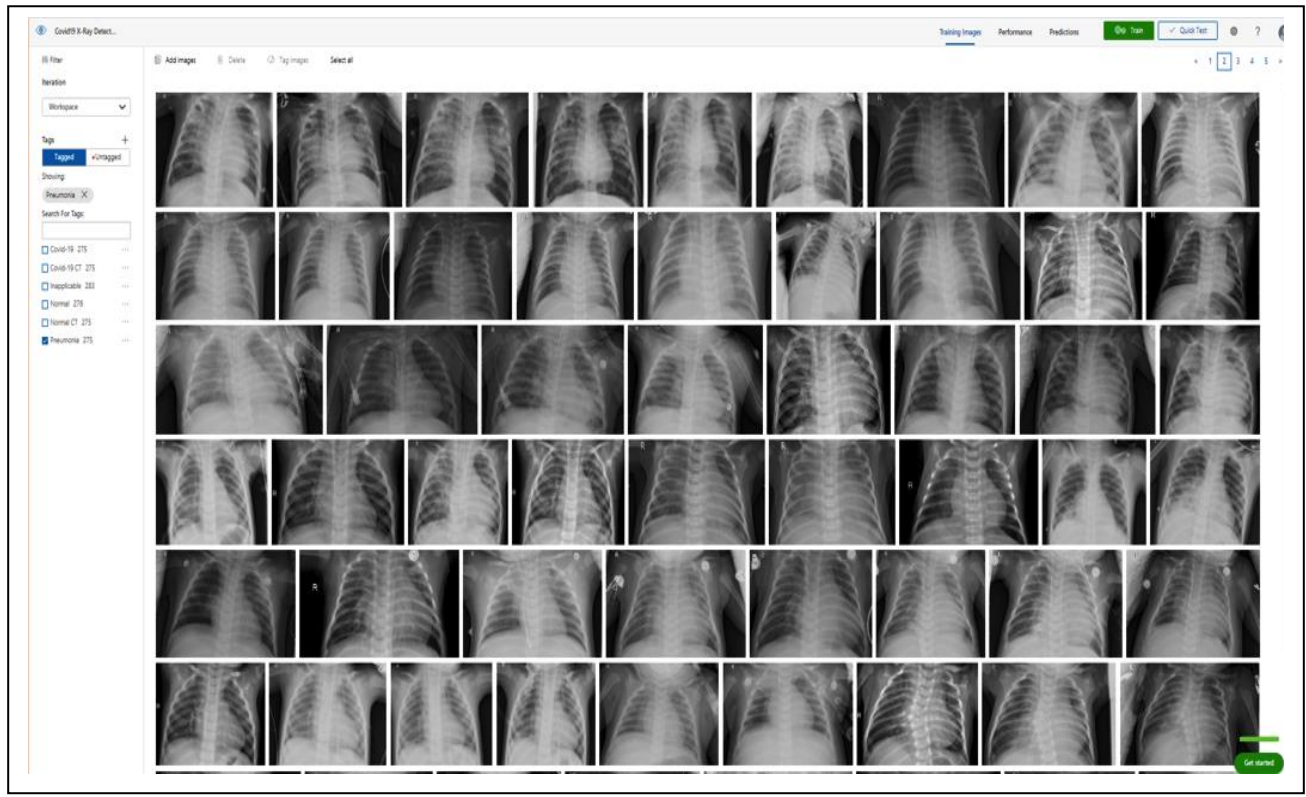

Fig 5: Pneumonia X-Ray Data loaded in Custom Vision

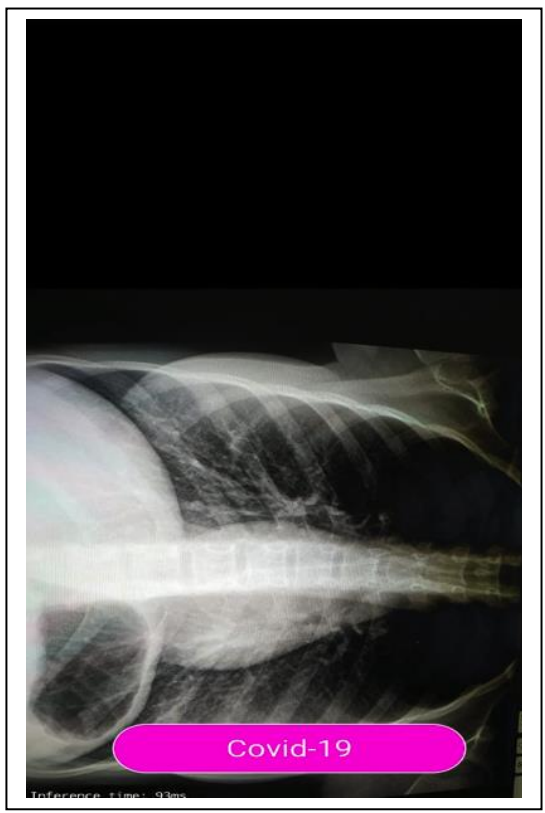 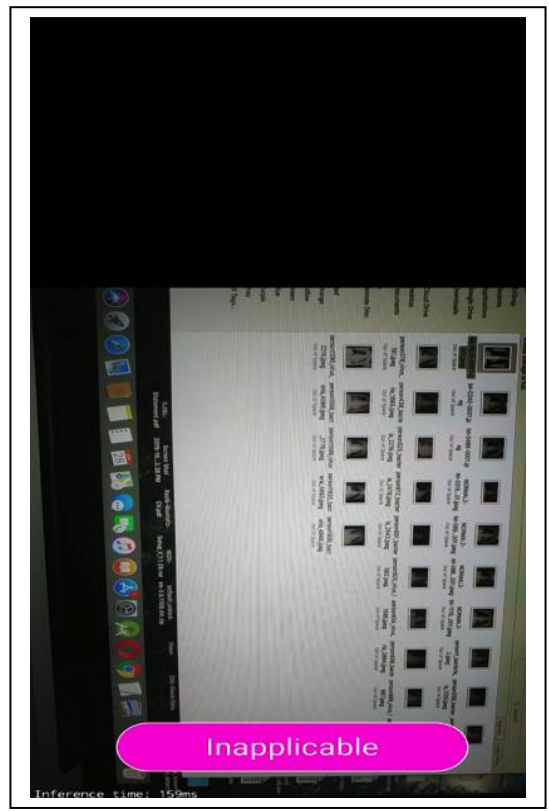

Fig 6: Smartphone App detecting COVID-19 X-Ray and Inapplicable images on Computer screen

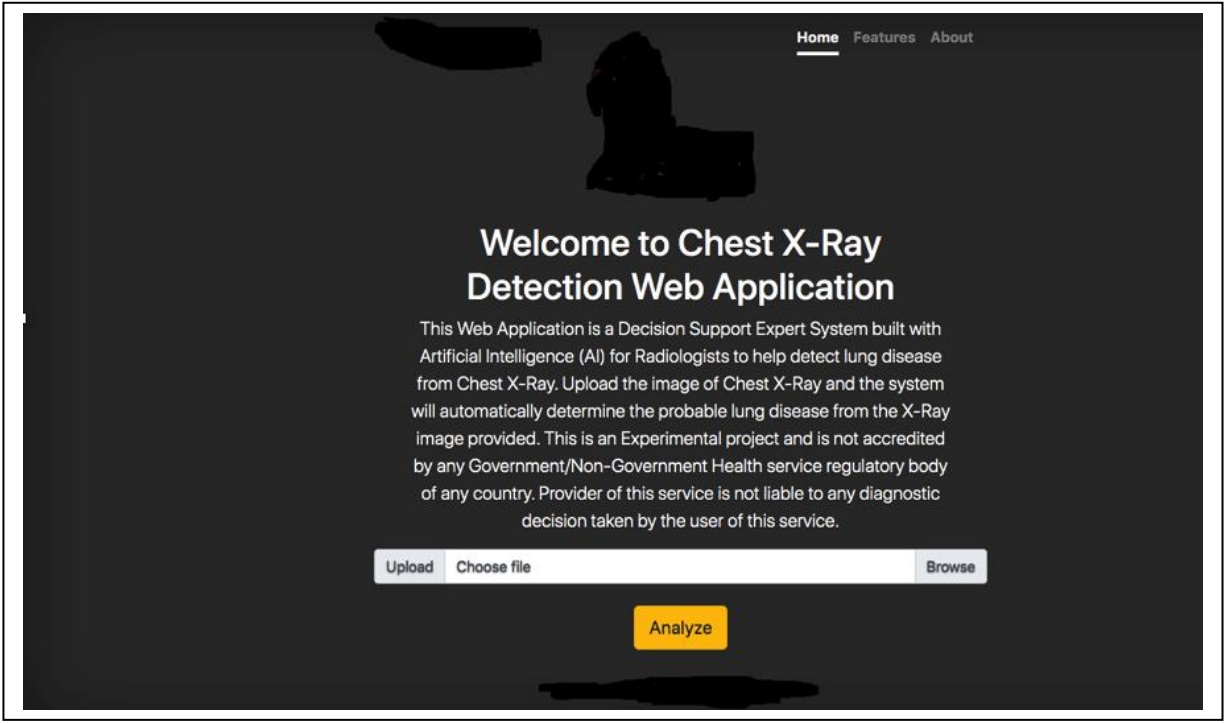

Fig 7: Web Application built with Custom Vision REST API

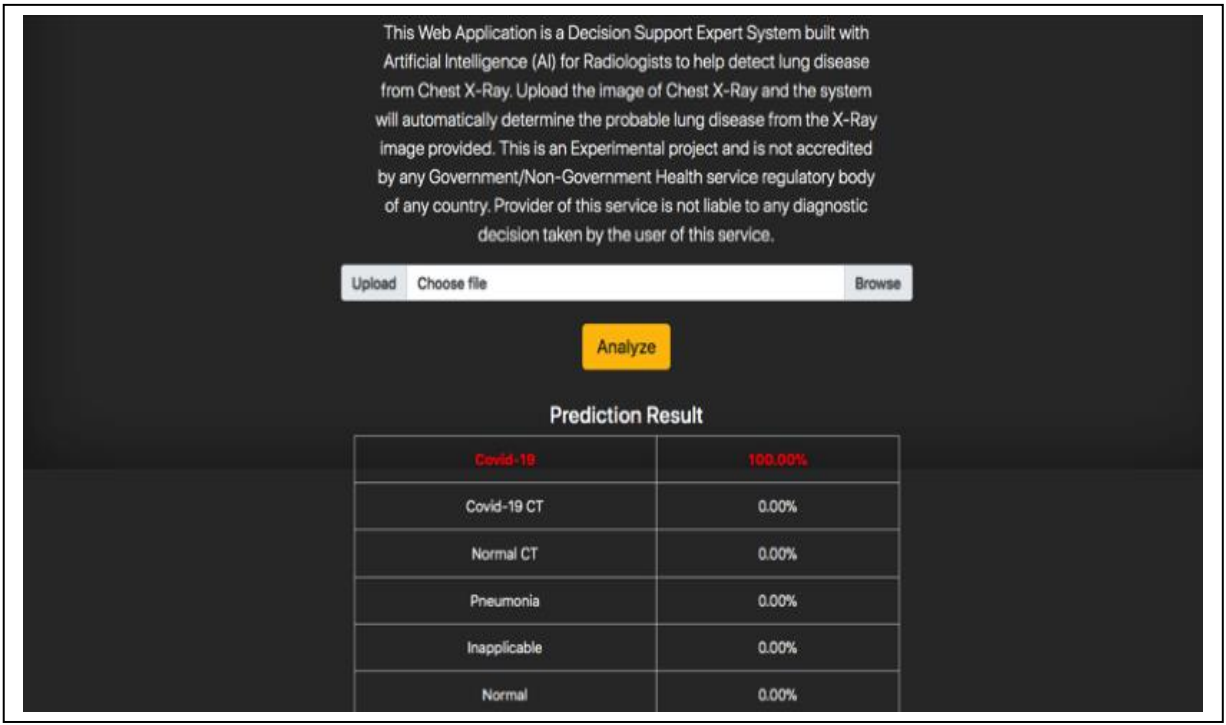

Fig 8: Web Application detected COVID-19 with 100% confidence after uploading an X-Ray

21